\documentclass[graybox]{svmult}

\usepackage{mathptmx}       
\usepackage{helvet}         
\usepackage{courier}        
\usepackage{type1cm}        
\usepackage{makeidx}         
\usepackage{graphicx}        
\usepackage{multicol}        
\usepackage[bottom]{footmisc}
\usepackage{url}
\usepackage{hyperref}
\usepackage{array}
\newcolumntype{C}[1]{>{\centering\let\newline\\\arraybackslash\hspace{0pt}}m{#1}}
\newcolumntype{L}[1]{>{\raggedright\let\newline\\\arraybackslash\hspace{0pt}}m{#1}}
\newcolumntype{R}[1]{>{\raggedleft\let\newline\\\arraybackslash\hspace{0pt}}m{#1}}
\usepackage{hhline}

\usepackage{floatrow}
\newfloatcommand{capbtabbox}{table}[][\FBwidth]
\graphicspath{{./plots/}}

\usepackage{blindtext}

\makeindex             

\begin{document}

\title*{The anatomy of Reddit: An overview of academic research}
\author{Alexey N. Medvedev, Renaud Lambiotte, Jean-Charles Delvenne}
\institute{Alexey N. Medvedev \at naXys, Universit\'{e} de Namur, ICTEAM, Universit\'{e} catholique de Louvain, Belgium  \email{an\_medvedev@yahoo.com}
\and Renaud Lambiotte \at Mathematical Institute, University of Oxford, UK \email{renaud.lambiotte@maths.ox.ac.uk}
\and Jean-Charles Delvenne \at  ICTEAM and CORE, Universit\'{e} catholique de Louvain, Belgium \email{jean-charles.delvenne@uclouvain.be}}
%
%
\maketitle

\abstract*{Online forums provide rich environments where users may post questions and comments about different topics. Understanding how people behave in online forums may shed light on the fundamental mechanisms by which collective thinking emerges in a group of individuals, but it  has also  important practical  applications, for instance to improve user experience, increase engagement or automatically identify bullying. Importantly, the datasets generated by the activity of the users are often openly available for researchers, in contrast to other sources of data in computational social science. In this survey, we map the main research directions that arose in recent years and focus primarily on the most popular platform, Reddit.  We  distinguish and categorise research depending on their focus on the posts or on the users and point to different types of methodologies to extract information from the structure and dynamics of the system.  We emphasize the diversity and richness of the research in terms of questions and methods and suggest future avenues of research. }

\abstract{Online forums provide rich environments where users may post questions and comments about different topics. Understanding how people behave in online forums may shed light on the fundamental mechanisms by which collective thinking emerges in a group of individuals, but it  has also  important practical  applications, for instance to improve user experience, increase engagement or automatically identify bullying. Importantly, the datasets generated by the activity of the users are often openly available for researchers, in contrast to other sources of data in computational social science. 
 In this survey, we map the main  research directions that arose in recent years and focus primarily on the most popular platform, Reddit.  We  distinguish and categorise research depending on their focus on the posts or on the users and point to different types of methodologies to extract information from the structure and dynamics of the system.  We emphasize the diversity and richness of the research in terms of questions and methods and suggest future avenues of research. }

\section{Introduction}
\label{sec:intro}

Understanding the dynamics and structure of human communication is a central research theme in computational social science. The increasing availability of digital traces of human interactions has allowed to quantify, at a large scale, a variety of phenomena. For instance, phone call logs led to the identification of the burstiness of human communication, typically organised into ``periods'' of short intensive communication followed by long periods of silence \cite{Karsai2011small}; Facebook and email data helped to confirm the smallness of the world i.e. the typical network distance between people is disproportionally small as compared to its size \cite{backstrom2012four}; tweet messages led to studies to uncover the mechanisms leading to information cascades \cite{zhao2015seismic}; etc. If early works initially focused on one-to-one communication, the emergence of new communication channels, such as Twitter or online forums, has opened the possibility to study collective discussions \cite{Aragon2017Review}. 

Collective discussions have not been invented by new media. As such, they have been and remain a major way for exchanging opinions and for producing collective decisions. Online forums provide a venue where Internet-goers post questions or comments, which may, or may not, trigger discussions from other members of the community. Understanding how people behave in online forums has important theoretical implications, to improve our understanding of collective thinking, but also practical applications, to improve user experience, increase engagement or facilitate the democratic process \cite{Aragon2016visualization}. The purpose of this Chapter is to provide an overview of the academic research on online discussion platforms, or online forums, and to bring together the variety of research questions considered the literature. Most of our attention is dedicated to the self-proclaimed ``front page of the Internet'' \cite{singer2014evolution} -- the website Reddit (\textsc{reddit.com}) -- which is the largest online discussion forum in the world as of today. Note that several other online discussion platforms have a similar architecture and have also been studied, for instance, in comparative studies; they include Digg, Hacker News, Slashdot, Epinions, Meneame, Barrapunto and even Wikipedia. 

The rest of this Chapter is organised as follows. Section \ref{sec:dataset} presents the datasets that can be extracted from Reddit and have been widely used by researchers. Academic studies are then divided according to their primary focus on the post or the users and are presented in Sections \ref{sec:posts} and \ref{sec:users} respectively. We conclude with a discussion and perspectives for future research.

\section{The Reddit dataset}\label{sec:dataset}

\begin{figure}[t]
\centering
\includegraphics[width=1\linewidth]{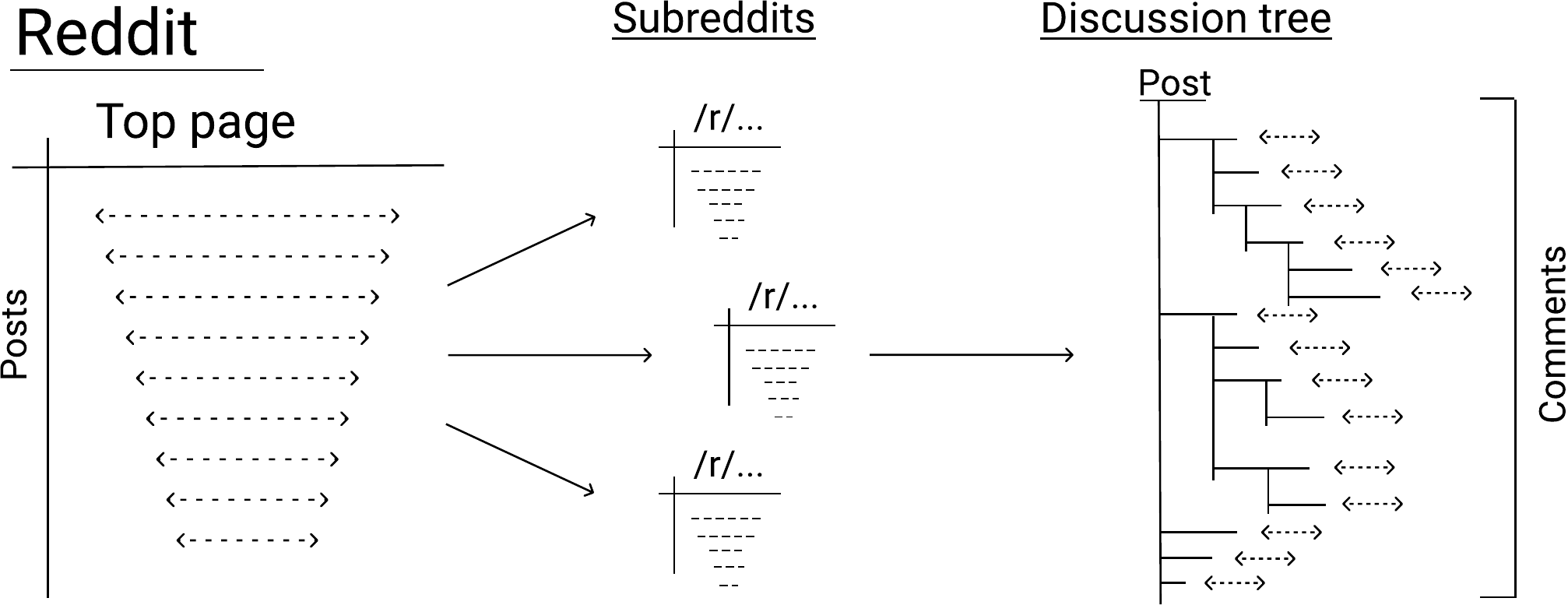}
\caption{The schematic structure of the Reddit platform.  The entry point is the top page of Reddit, which is fed of posts from the subreddits followed by a registered user (or from all subreddits, for an anonymous user) and ranked according to votes and posts' age. The user may further proceed to top page of a specific subreddit, where the feed narrows down to only posts from a chosen subreddit. Each post can be upvoted or downvoted and has an attached section of comments. The comments are structured as a rooted tree by the reply-to relation to other comments or the post itself. }
\label{fig:reddit_scheme}
\end{figure}

Reddit (launched in 2005) is a social news aggregation, web content rating and discussion website, ranked as \#6 most visited website in the world with 234 million unique users (as of February 2018)\footnote{\url{https://en.wikipedia.org/wiki/Reddit}}. A schematic structure of Reddit is illustrated in Figure \ref{fig:reddit_scheme}. Registered users submit posts that contain a title, an external link or a self-written piece of content, which immediately become available to the  whole audience of Reddit for voting and commenting. The voting system permits only registered users to upvote (give a positive +1 vote) or to downvote (give a negative -1 vote) on posts and comments. Comments form a \textit{discussion tree}, which can be described as a rooted tree, where the root is a designated node representing the post itself and each other node represents a comment. There is a link between two nodes if there is a `reply-to' relation between them. 

The huge posting space of Reddit is divided into subreddits -- \textit{self-created} communities of users, united by a certain topic. Every submitted post has subreddit name as an intangible attribute. Each subreddit and Reddit itself has a so-called ``top page'' -- the feed where post titles with voting and commenting links are delivered to users. Two factors influence the post's ranking position there: 1) time and 2) voting \textit{score}, or otherwise called \textit{karma}, which is basically the difference between upvotes and downvotes. High score posts have a higher chance of appearing at the top page. However with time, newer information replaces the older in the feed. Users can follow subreddits, but not other users, which constitutes the main distinction with social network platforms, like Facebook or Twitter, where users follow a person and not a content. Other platforms have a similar structure. For example, Slashdot (launched in 1997) is made of news stories, together with comments moderated by selected users, but not by an open voting system\footnote{\url{https://en.wikipedia.org/wiki/Slashdot}}. Only a fixed number of topic-based subsections is available. Hacker News (launched in 2007) is an online community very similar to Reddit but with only two pre-made topic subsections (subreddits)\footnote{\url{https://en.wikipedia.org/wiki/Hacker_News}}. Digg (launched in 2004) acts currently as a news aggregator, but it formerly was a socially curated platform with a post submission, commenting and voting system like Reddit\footnote{\url{https://en.wikipedia.org/wiki/Digg}}. Meneame is a Spanish analogue of Digg, Barrapunto is also a Spanish version of Slashdot \cite{Gomez2011}. 

\begin{figure}[t]
\centering
\includegraphics[width=1\linewidth]{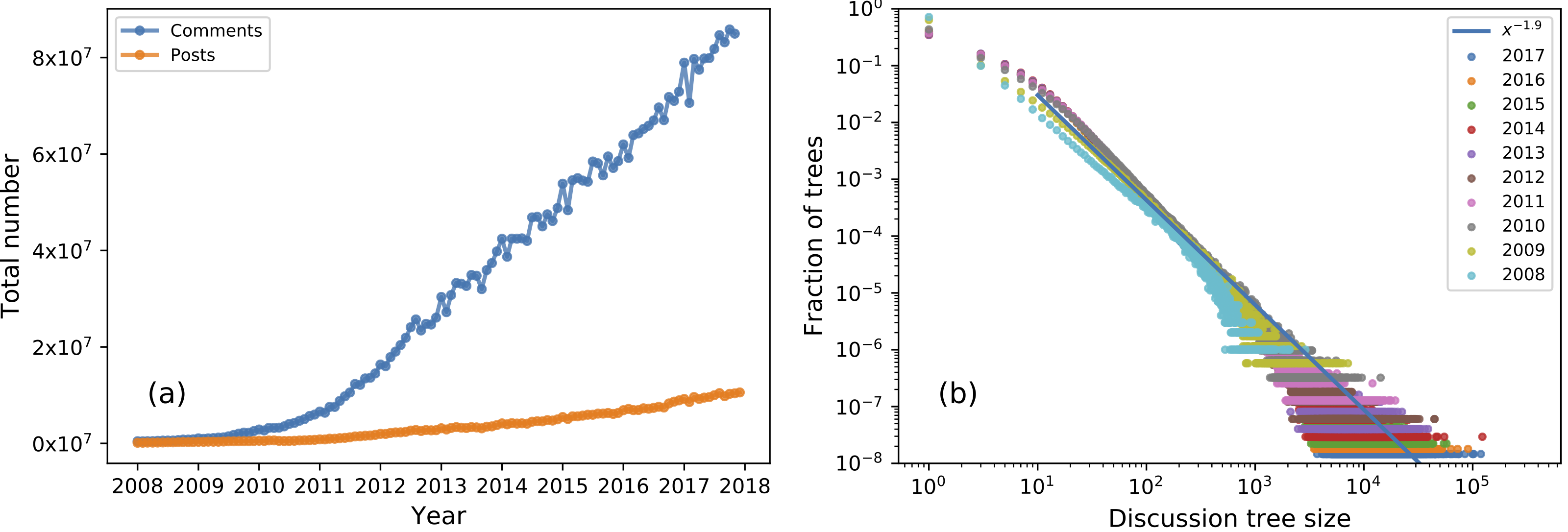}
\caption{The evolution of Reddit from Jan 2008 till Jan 2018: a) monthly counts of posts and comments, b) distribution of discussion sizes. One may notice an exponential increase in the activity counts, but the discussion size distribution follows a similar shape, close to the power-law with exponent $\alpha$ ranging from around $1.7$ for early years to $1.9$ for later.}
\label{fig:reddit_presentation}
\end{figure}

We chose to dedicate our attention to Reddit because of the variety of ways the system is self-organized. As it is mentioned later in the chapter, such self-organization provides place for free movement, social herding, organised attacks, trolling, etc. Subreddits as topically self-identified entities create real communities in virtual space, with their own rules, language features, intrinsic rules, jokes, etc. Such annotated text corpus becomes valuable for training machine learning algorithms \cite{NormanAI2018}. General availability of information brings in different dynamics of information spread in comparison to direct followers social media (e.g. Twitter), the volume of information poses the problem of missed content, or the posts that potentially could gain comments or votes, but was missed in the avalanche of other posts. Reddit allows sharing a wide variety of content. One may use Reddit data to uncover the relationships between different Internet services.

Reddit has gained a central place in the scientific literature thanks to the openness, richness and quality of its data, which allows to perform longitudinal studies of the whole system and, critically, to ensure reproducibility of the results.  Jason Baumgartner, under the Reddit name Stuck\_In\_The\_Matrix, did a tremendous amount of work when attempted to collect a full dataset of posts and comments, going back to the creation of the site \cite{dataset_reddit}. The figures of this Chapter have all been prepared from this dataset. For instance, basic numbers on the growth of the site and the total sizes of discussions are found in Figure \ref{fig:reddit_presentation}. His data repository also contains the data from the platform Hacker News \cite{dataset_link}. 

Despite its recognized quality, one should be careful while using the dataset for research purposes. Gaffney and Matias \cite{Gaffney2018} report several inconsistencies in the data. Their approach is based on the fact that a post or comment id is essentially an integer number in base36 format, thus, allegedly, all the continuous range of numbers must be present in the dataset. For example, comment and post data before 2008 appears to be hugely corrupted, having around 80\% of posts missing, as well as 90\% posts information from few months data at the interface between 2009 and 2010. In total, across the time interval Jan 2006 and Feb 2016, the authors report 0.043\% missing comments and 0.65\% missing posts. 

The risks of missampled data obviously cause commenting/posting rate distortions, missing information in user time series and possible inability to reconstruct certain discussion trees. Nevertheless the data from Jan 2008 and later is fairly consistent for detailed look and especially in large scale studies the inconsistencies may be safely disregarded due to their smallness. The system has sustained exponential growth \cite{singer2014evolution}, thus data volume in early years is negligibly small compared to today's numbers. The proposed approach to measure missing data also raises a question of applicability, since the gaps in consecutive numbering may be due to inner technical features of the website. This may be supported by the fact that a newly published rescraped data contains the same missing values found in the data before \cite{dataset_rescraped}. One may use directly the Reddit API for consistency checks\footnote{https://praw.readthedocs.io/en/latest}.

Although missing data causes reasonable inconsistencies, the present data has a few important peculiarities to consider. There exist posts and comments from authors, whose accounts were deleted and the author name of the comment turns to default name ``[deleted]". According to the letter appendix to \cite{Gaffney2018} such comments comprise around 25\% of the data. This fact imposes a greater obstacle on studies of user participation in discussions and reply networks. Another problem comes with posts or comments, that were deleted by user and removed due to moderation. In the current version of the API the text body of such comments and posts is correspondingly marked as ``deleted'' and ``removed", but it is not clear what happened in early years.

\section{From the perspective of posts}\label{sec:posts}

Posts are at the heart of the platform structure and dynamics. Once posted, they may gain attention and receive feedback in the form of votes and comments, thereby obtaining a good ranking and even more attention. They may also go quickly unnoticed in the avalanche of newer posts. The next subsection is dedicated to the topic of popularity prediction in online platforms. This longstanding problem has been studied
in various online systems. As first step, we thus provide a quick overview of research on the wide spectrum of online systems before concentrating on Reddit specifically.

A standard ingredient in predictive models is the incorporation of quantitative features that tends to correlated with the popularity of posts. Features can be structural, dynamical, textual and even associated to the author of the post. Initial works proposed simple statistical models based on regression, Poisson or Cox processes, but with the developments of machine learning, more elaborate methods based on neural networks have emerged.

\textbf{Popularity prediction.} Anyone who has ever used online social networks is familiar with the concept of ``likes'' and ``dislikes'' -- the way of expressing attitude towards a piece of content on a  binary scale. If   ``page views'' have long been the dominant measure of the success of a content, more and more platforms have moved to voting systems where the number of positive votes is the measure of  popularity. Discussion platforms use  systems of upvotes and downvotes for different purposes, ranging from the automatic discovery of appreciated items and its delivery to a wider audience, to the moderation of discussions to protect from spam or malicious content. For these reasons, good models of popularity prediction are  of interest for both content creators and platform curators.

In general, the problem of popularity prediction has been considered in various online social systems. Early studies, e.g. in YouTube and Digg,  found a direct relation between content's initial popularity (in terms of views and upvotes) and its future counts \cite{Szabo2010}, but more sophisticated models have been proposed since then. For instance, Lee et al. \cite{Lee2012} have modelled the lifetime of discussions on Myspace with the Cox proportional hazard regression model. They selected the number of ``risk'' factors, fitted from the data for each thread, which were further used as a predictor of threads hitting a threshold number of comments. Mishne and Glance \cite{Mishne2006} analysed the corpus of comments in weblogs and the relation between weblog popularity and commenting patterns in it. Tsagkias et al. \cite{Tsagkias2009} analyse the corpus of comments under news stories in regional Internet news agents. The authors propose a model that predicts the commenting popularity prior to article publication in two cases: first, if there is a potential to receive comments and second, if the article receives ``low'' or ``high'' comment volume. Bandari et al. \cite{Bandari2012} also investigated if popularity of news articles can be  estimated  even before their posting online. 


\begin{figure}[t]
\centering
\begin{minipage}[b]{0.48\linewidth}
\centering
\includegraphics[width=0.8\linewidth]{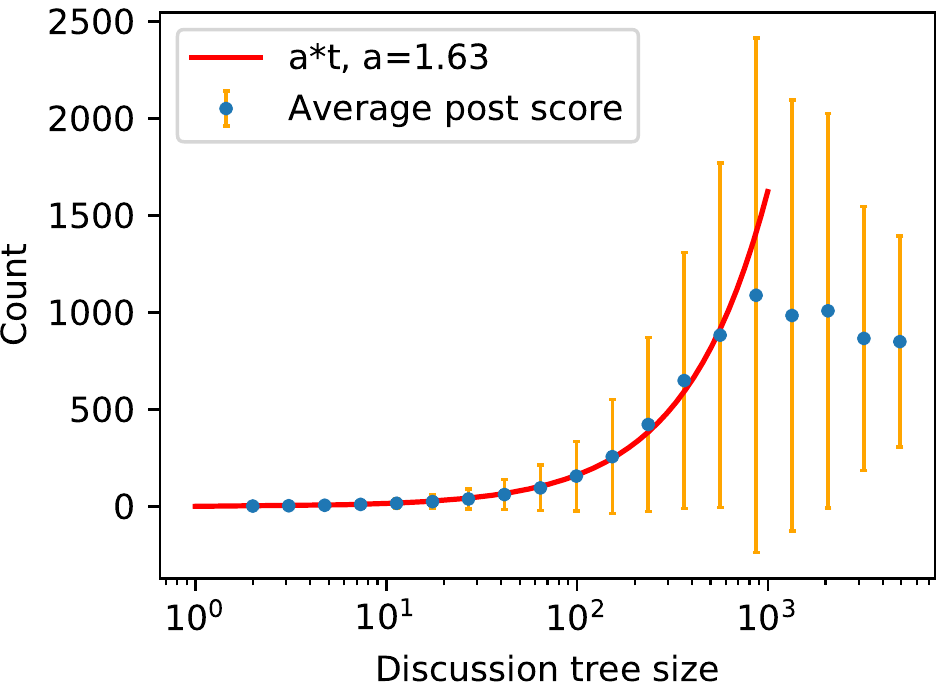}
\end{minipage}
\begin{minipage}[b]{0.48\linewidth}
\centering
\includegraphics[width=0.8\linewidth]{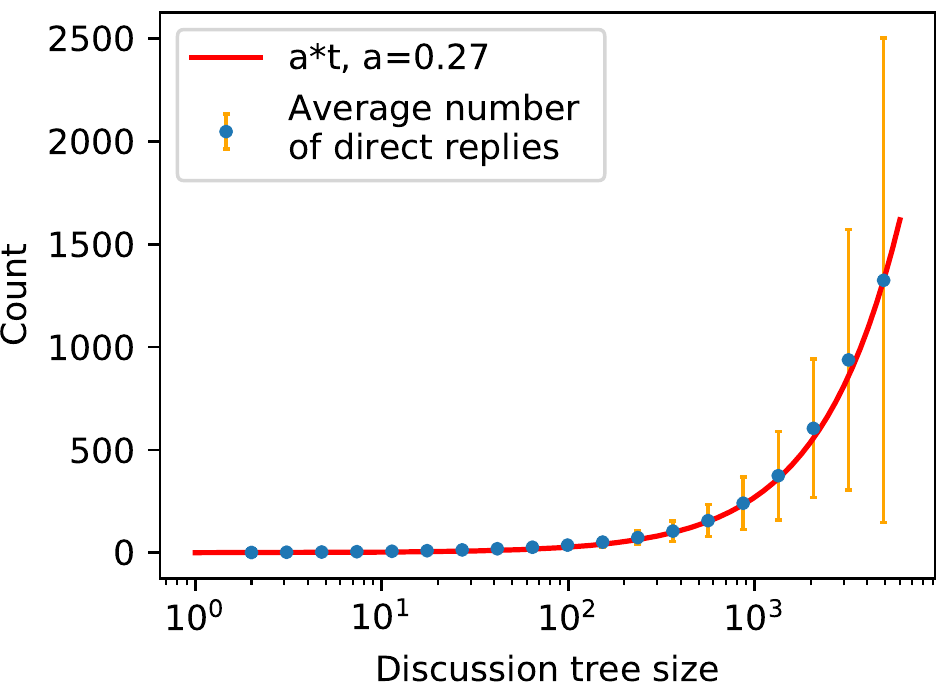}
\end{minipage}
\caption{Average posts' score versus discussion size (left figure) and number of direct replies to a post versus discussion size (right figure) in Reddit. The data shows that average values may be fitted with a linear trend up until a certain value. Dots represent average counts, bars show standard deviation. The figures are based on data from year 2009, but other years show similar results.}
\label{fig:score_degree_size}
\end{figure}

The Reddit dataset shows a proportional relation between the score of a post and the size of its discussion tree \textit{on average}, as shown in Figure \ref{fig:score_degree_size}. This may shed some light on general aspects of the posts' popularity, however, in each particular case, it is more important to make a more tailored estimation of submissions' score. A number of works has been dedicated to the prediction of scores on Reddit. Horne et al. \cite{Horne2017} found a number of textual and temporal features of high score comments by considering the discussion threads from 11 popular subreddits appearing in a 6 month period of 2013. The authors proposed a machine learning model for predicting comments' score and pointed the differences in users' preferences in subreddits. In particular, they claimed that timing of the comment, its relevancy and novelty have positive impact, but  stale memes or high user ranking (overall number of positive comments in a user's history) does not affect or even pushes down the average comment score. It was observed that moderation does not always impact proper behavior in the community and may shorten the life of a discussion thread. 

Recurrent neural networks (RNN) have also been employed to measure community endorsement. Fang et al. \cite{Fang2016} constructed an RNN trained to predict comment scores. Instead of controlling for the submission context, the model learns latent modes of submission context and examines how the  context  relates to different levels of community endorsement. On a dataset of three popular subreddits, they achieve a good performance, on average, and show that high score comments are usually harder to predict than lower ones. High scoring comments tend to be submitted early in the discussion and the number of direct replies is not smaller than the height of its hanging discussion subtree. Low and medium score comments have a number of direct replies less than the height of  a discussion subtree, indicating the presence of a further discussion. Low score comments tend to come later in the discussion overall, but also later in terms of the group of responses to a parent comment.

While structural features have been shown to be good predictors, researchers have started working on extracting textual linguistic features in order to gain predictive power. In this direction, Jaech et al. \cite{Jaech2015} have reported general improvement of machine learning classifiers in the problem of ranking comments that appear in a fixed time window in a discussion thread. Note, however, that the gain was reported to be marginal. Later, Zayats and Ostendorf \cite{Zayats2018} constructed a specific type of RNN called LSTM (long short term memory) for the same purpose of  predicting comment scores. The proposed model uses structural and temporal comment features, as well as textual linguistic features of the comments. The authors achieved a slightly better performance (increase in average F1 score from 50 to 54 on average) on a dataset of three subreddits studied earlier in \cite{Fang2016}. They found that controversial comments (that further generate a wide discussion in terms of a discussion tree) tend to be overpredicted (with lower score than predicted) and jokes and funny comments, on the contrary, were mostly underpredicted (with a higher score than predicted). Linguistic context was found to be helpful in prediction tasks and  words of underpredicted comments were aligned with comments of positive score, but words associated with overpredicted comments did not show any significant correlation. In another work, Hessel et al. \cite{Hessel2017} considered pairs of posts, submitted within a very short time interval into the same communities (to exclude  timing bias) and predicted the more popular posts in those pairs. Six primarily image sharing subreddits were selected for the study, with a set of features including textual and temporal information, but also image features assessed by the deep neural networks. The authors  concluded that  user-centric characteristics, e.g. previous popular submissions, and content-specific features, e.g. more complicated images and simpler titles, make a good predictor of popularity of the submission. The authors also reported an accuracy comparable to that of human classification.

\begin{table}[t]
\begin{tabular}{ L{2.3cm} | C{4.2cm} | C{2.3cm} | R{2.3cm} } 
 Article & Task & Dataset & Methods \\ 
\hhline{====} 
Horne et al. \cite{Horne2017} & Predict high scoring comments, assess the impact of thread moderation & Reddit dataset \cite{dataset_reddit}, 11 top subreddits & Linear regression, sentiment analysis \\ \hline
Fang et al. \cite{Fang2016} & Predict final score of comments & Reddit, three chosen subreddits & Recurrent neural networks (RNN) \\ \hline
Zayats, Ostendorf \cite{Zayats2018} & Predict final score of comments & Reddit, three chosen subreddits & RNN with long short term memory (LSTM) \\ \hline
Hessel et al. \cite{Hessel2017}  & Given a pair of submissions, predict the one with higher final score  & Reddit dataset \cite{dataset_reddit}, 6 image-sharing subreddits & Image description (convolutional neural networks), LSTM \\ \hline
Stoddard \cite{Stoddard2015}  & Determine inherent quality of posts and to predict high-scoring posts & Hacker News; Reddit dataset \cite{dataset_reddit}, 5 top subreddits & Poisson processes \\ \hline
Lakkaraju et al. \cite{Lakkaraju2013}  & Predict popularity of resubmitted content & Reddit, unique dataset of resubmitted images & Poisson regression \\ \hline
Arag\'{o}n et al. \cite{Aragon2017} & Review of the models of discussion trees & Reddit, Slashdot, Meneame, Barrapunto, etc. & Review \\ \hline
Medvedev et al. \cite{Medvedev2018} & Model structure and predict dynamics of discussion trees & Reddit, dataset \cite{dataset_reddit} & Stochastic Hawkes processes \\ \hline
\end{tabular} 
\caption{Short summary of the articles with studies on Reddit, presented in Section~\ref{sec:posts}}
\end{table}

The content popularity is influenced by various factors and public endorsement may not properly reflect its inherent quality \cite{Sinatra2018}. This phenomenon has been studied by Stoddard \cite{Stoddard2015} by means of a Poisson regression model that infers the intrinsic quality of posts from  voting activity of the users. The author collected a unique dataset of users' voting time series by tracking a number of top posts on front pages of 5 subreddits and the front page of Hacker News, and used data to predict final score of posts. A variable of quality was then introduced in the model parameters and was shown to correlate with the total post scores, although there were several situations when similar quality posts had different scores and vice versa. Amongst others, the mechanism of making popular more visible than less popular ones leads to a multiplicative process that increases the variance of popularity, with the effect of making a substantial fraction of the posts ignored. According to Gilbert \cite{Gilbert2013widespread}, Reddit overlooks 52\% of the most popular links the first time they were submitted. Lakkaraju, McAuley and Leskovec \cite{Lakkaraju2013} explored this idea and showed that resubmissions of the same piece of content may gain more popularity than an original submission. The authors collected a dataset of image submissions between 2008 and 2013, where each image was resubmitted roughly 7.9 times\footnote{The authors used \url{karmadecay.com} -- the reverse image search tool specifically designed for Reddit.}. Same pictures can in principle be resubmitted to different subreddits, thus the authors employed a success metric compared to the average post score in the community. The authors also proposed a statistical model that predicts the expected score of a resubmitted picture. The model parameters include the inherent content popularity, penalties from previous success and previous submissions to other communities or to the same community twice. Overall, the study supports the hypothesis that a high quality content `speaks for itself' and determines its score. The choice of  subreddit plays an important role -- the model shows that the content, resubmitted to the same subreddit, in general was unlikely to be popular, as well as whether  the content was previously highly rated in a popular subreddit (with a high number of visitors or subscribers). This effect gradually disappears with time, indicating forgetfulness of the audience. Titles of resubmissions were also found indicative: if the title is novel, written using subreddit-specific words and sentiment orientation, the submission has higher chances to receive positive feedback. Similarly, Glenski et al. \cite{Glenski2017, GlenskiWeninger2017}  found that users mostly vote on posts only after glancing at the title, without proper reading of the content or the discussion. The authors of \cite{Lakkaraju2013}  also performed an \textit{in situ} experiment: they manually chose and resubmitted 85 images from the dataset, select a ``good'' and a ``bad'' title according to the model for each picture and post them in two different subreddits. The post scores, gathered after one day, show that submissions with a ``good'' title generated scores three times higher  than the ``bad'' ones. 

\begin{figure}[t]
\centering
\includegraphics[width=1\linewidth]{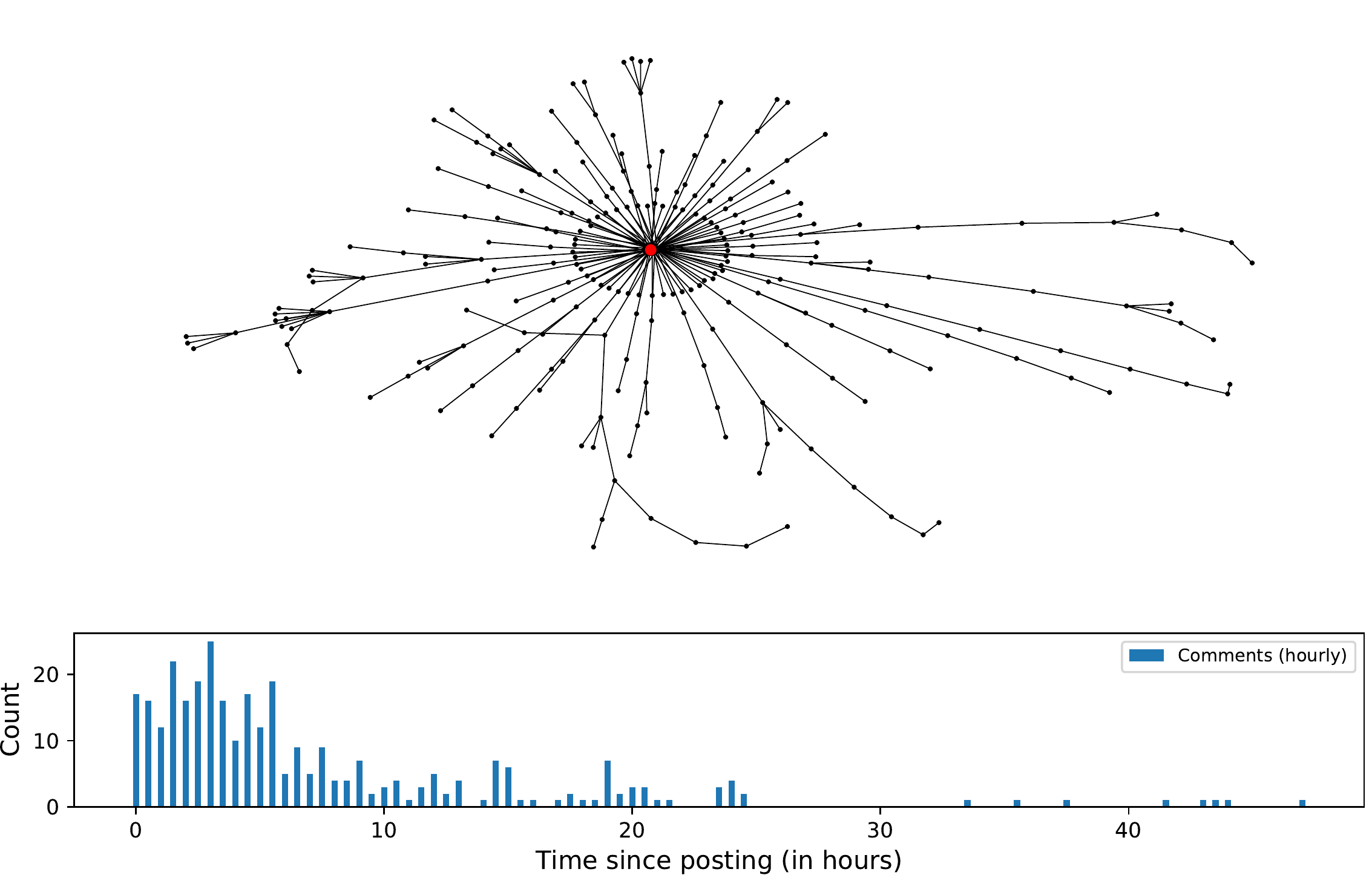}
\caption{Sample discussion tree of a post on Reddit with a histogram of comment arrival. Central large red node depicts the post, comments depicted in  black. The histogram presents hourly aggregation of comments arrival. }
\label{fig:sample_post_activity_timings}
\end{figure}

\textbf{Generative models for discussion trees.} As mentioned above, comments under the post form a rooted tree. Such trees have dynamic nature and their temporal growth reveals the dynamic of attention to the post (see an example of a tree and a histogram of comment arrival on Figure \ref{fig:sample_post_activity_timings}). Generative models for discussion trees mostly question the tree structure of a discussion while disregarding the comments' or posts' textual features and exact timings. The reader may refer to the extensive review on generative models given by Arag\'{o}n et al. \cite{Aragon2017Review}, and here we give only a brief overview of the representative contributions. 

Gomez et al. \cite{Gomez2011} considered discussion trees in four large Internet boards (Slashdot, Barrapunto, Meneame and Wikipedia) and proposed a generating model based on preferential attachment mechanism (PA model) with respect to the comment degree and the root bias. Later Gomez et al. \cite{GomezLitvak2013} enriched this model by incorporating a notion of novelty of comments, which is represented by an exponentially decaying function of attractiveness. The model showed better results in likelihood of representing the tree structure and reproduced well the width/depth relation for discussion trees. Lumbreras et al. \cite{Lumbreras2017} proposed to enrich the PA model with the notion of \textit{roles}, which are latent functions of community members. The PA model defines a set of parameters that regulate the place of attachment of a new comment. The authors suggest that this set of parameters is different for different users and propose to group them into the role sets, which are inferred accordingly. Despite of this extra natural assumption, the gain in model likelihood is marginal.

Arag\'{o}n et al. \cite{Aragon2017} considered the social system Meneame, where the change in discussion representation happened in 2015 from a plain list to a structured threaded view. This change was observed to have an impact on the structure of discussion trees and the authors further enrich the PA model, with a reciprocity term that captures the tendency of posting authors to reply back in the discussion. It was observed that change of the platform interface had a positive effect on reciprocity, as well as on other parameters in general. 

The above mentioned models focus exclusively on the structure of discussion trees while leaving out the continuous time dynamics of the comment attraction process. Kaltenbrunner et al. \cite{Kaltenbrunner2007} found that comment arrival time in Slashdot discussions fits well by double lognormal distribution, although the fitting quality depends on a circadian rhythm of the site. Based on this finding, the authors propose the prediction model, that predicts the total number of comments in a discussion thread. The dynamical aspect of tree generation was first studied by Wang et al. \cite{WangHuberman2012}, where the authors introduce a merely theoretical model for the structural and temporal evolution of discussions. The temporal evolution was described as a L\'evy process with power-law interevent time distribution, when newly arriving comments were assumed to attach to the existing tree under the simple PA rule. The model was inspired by  empirical observations of such discussion boards, like Reddit, Digg and Epinions, however the mean-field nature of it limits its calibration with real-world datasets. Medvedev et al. \cite{Medvedev2018} used a Hawkes process along with its branching tree interpretation to jointly model structure and dynamics of discussion trees in Reddit. The model was further used for prediction of the discussion flow, performing better than contemporary models of cascade dynamics.

\begin{figure}[t]
\centering
\begin{minipage}[b]{0.48\linewidth}
\centering
\includegraphics[width=0.8\linewidth]{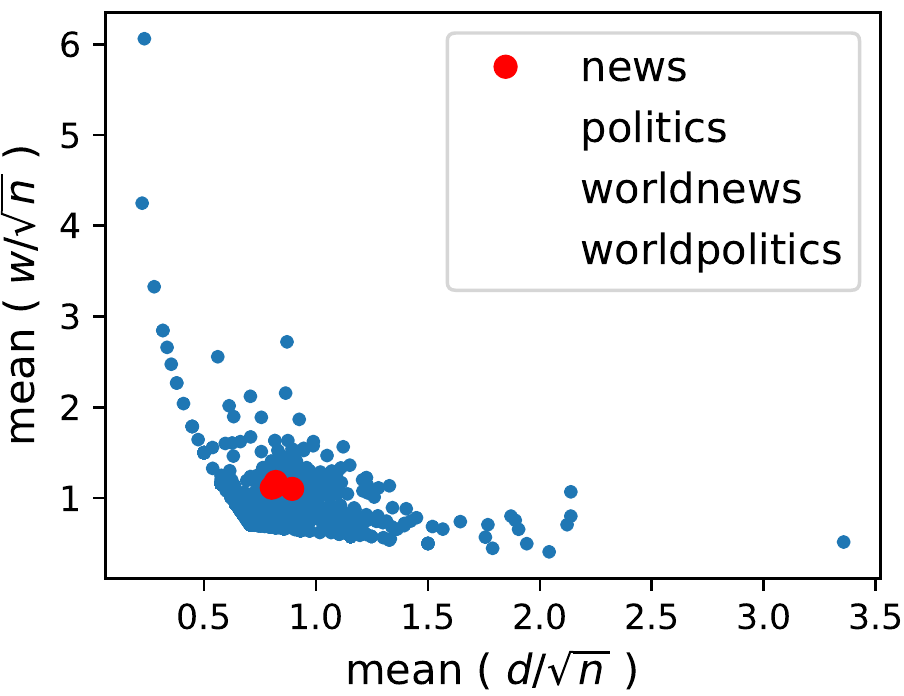}
\end{minipage}
\begin{minipage}[b]{0.48\linewidth}
\centering
\includegraphics[width=0.8\linewidth]{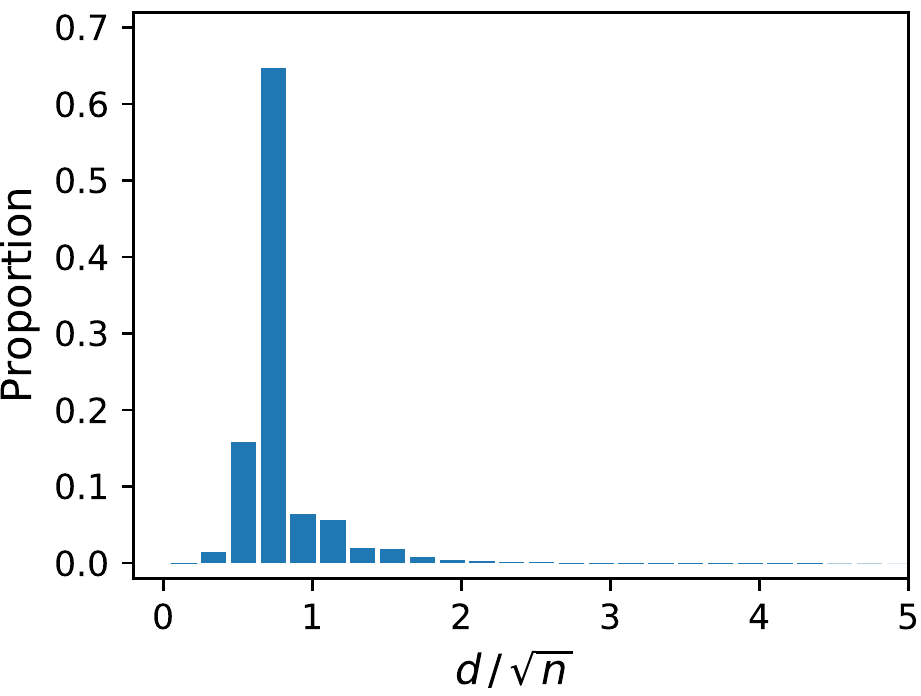}
\end{minipage}
\caption{Average scaled discussion tree depth $d/\sqrt{n}$ versus discussion tree width $w/\sqrt{n}$, where $n$ denotes scaling by tree size (left figure) and scaled depth distribution (right figure) in Reddit. Large red dots on the left figure denote political and news subreddit. The figures are based on data from year 2008.}
\label{fig:depth_width}
\end{figure}

The discussion trees in Reddit exhibit interesting peculiarities in their structure. For instance, the trees exhibit the so-called ``root bias'', which means that regardless of the tree size the degree of a root is on average larger and more broadly distributed than the degree of a comment. This may arise from the fact that the direct comment is induced by the post itself, not the subsequent discussion, and that such discussion is only shown after opening a specific link under the post. Root bias is not a unique feature of Reddit, other platform like Slashdot, Barrapunto, Meneame, Wikipedia also own it \cite{Gomez2011, GomezLitvak2013}. Analysis of political discussion trees \cite{GonzalezBailon2010} and reply trees in Twitter \cite{nishi2016reply} suggest considering the width/depth relation for the trees. For example, reply trees in Twitter were shown to have a duality of being either long chain-like trees (low width, large depth) or star-like trees (vice versa). It is a commonly known fact that critical branching trees of size $n$ have depth $d$ and width $w$ proportional to $\sqrt{n}$ \cite{marckert2003depth}, which turns out to be the case for the Reddit discussion trees, where the scaled value of $d/\sqrt{n}$ is well-centred (see Figure \ref{fig:depth_width}). This is also supported by the fact that the branching tree model reproduced the trees better than the PA model \cite{Medvedev2018}, which apparently has not more than logarithmic depth \cite{cohen2003scale, dommers2010diameters}. Political discussion trees on Slashdot were shown to have larger depth and width in comparison to other trees. Discussions in political subreddits also show large depth and width, but apparently this is mainly due to the fact that on average it has larger participation rate, which is clear from the mean scaled width and depth values for different subreddits on Figure \ref{fig:depth_width}.

\textbf{Other.}  Zannettou et al. \cite{Zannettou2018} perform a study of meme evolution and propagation across different platforms, e.g. Twitter, Reddit, 4chan and Gab, where the last two are image boards structurally similar to Reddit. The authors' analysis helps to reveal the influential actors in meme ecosystem, both in terms of creation and propagation, the authors build clusters of similar memes and make an analysis of reciprocal influence between the observed communities using Hawkes processes. 

\section{From the perspective of users}\label{sec:users}

So far we emphasised the posts as the central pieces of information driving the dynamics of the platform. We now focus on the person that hides behind each post, comment, like or dislike, and review studies on the behavioral features of users. 
The main approaches in this section are observational and data-oriented. Statistical methods are employed in order to analyse the users track records and their community organisation is uncovered by analyzing the  network of relations
between actors.

\textbf{Activity patterns.} Observing the actions of users on a website can lead to interesting conclusions. Glenski et al. \cite{Glenski2017, GlenskiWeninger2017} have studied a dataset with all recorded activity of 309 Reddit users within one year. The activity log included the information on all clicks, pageloads and votes made within the \textsc{reddit.com} domain. As  expected, the majority of users prefer passive browsing and rarely interact with the content (only 16\% of users produce more than 50\% of interactions). Users mostly vote on posts on average after only browsing the title (73 \% of posts), although a non-negligible fraction (17\%) of participants follow the link of the post and browse the section of comments before giving a vote. It was noted that users' probability of interaction with a given post decreases with the ranking of the post on the top page of Reddit as well as on subreddits. Text analysis of post titles shows that the probability of interaction increases with the reading ease of the title, i.e. as they use shorter words, and smaller sentences. The authors used the concept of activity sessions, which are the periods of user activity starting from an interaction and finishing after 1 hour without consecutive interactions. This terminology and a 1-hour threshold were adopted from Halfhaker et al. \cite{Halfaker2015} and Singer et al. \cite{Singer2016}. The authors reported a mean session length of  53 minutes. However,  most participants had  much shorter sessions prevailing ($>$3 mins). Singer et al. \cite{Singer2016} also studied performance deterioration within sessions of active commenting on Reddit, where sessions of increasing intensity, i.e. how many posts users produced during sessions, are associated with the production of shorter, progressively less complex comments, which receive declining score. In this work, the authors found a similar prevalence of short sessions and sessions, presented a daily circadian rhythms. 

\begin{figure}[t]
\centering
\includegraphics[width=0.9\linewidth]{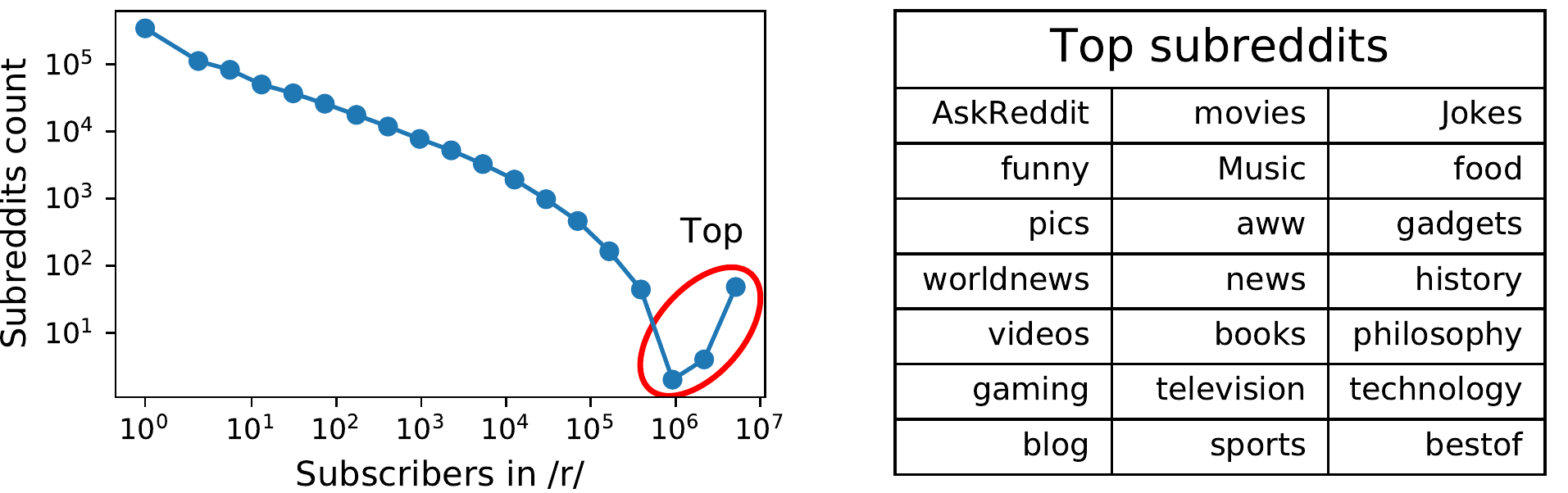}
\caption{Distribution of the number of subscribers of subreddits. The largest subreddits in the rising tail of this distribution are shown along in the table.  }
\label{fig:subreddits_subscribers}
\end{figure}

\textbf{Community loyalty.} While some of the previous results also applied to other online platforms, this section is devoted specifically to Reddit, a platform where the content is by default submitted to thematic communities, or subreddits. These subreddits are in principle open to anyone, but users can follow particular subreddits of their interest in order to customizing their feed. Interestingly, subreddit sizes are close to have heavy-tailed distribution and there exists a fraction of subreddits-outliers with a huge number of subscribers \ref{fig:subreddits_subscribers}. Looking closer at those we find that many of these ``top'' subreddits are those which are proposed to sign up for by default at the registration of a new user. 

Tan and Lee \cite{Tan2015} studied the posts of users across the subreddits and found that users on average tend to explore and continuously post in new communities, moreover they tend over time to share their activity evenly between a small number of communities with diverse interests. Differences in posting patterns of users may be used for prediction of the users' future settlement status in a community. Vagrant users, on average, post to more similar communities in comparison with the settling users, they use different language patterns from those existing in a community and their posts receive less attention in terms of score. Score of the first post may generally act as a predictor of further postings. The posting activity rate alone showed to be a bad predictor of the future settlement status in a community. An interesting finding is that the very same users tend to use different vocabulary when posting in different communities, therefore adapting to the community language. Hamilton et al. \cite{Hamilton2017} defined loyal communities as the ones that retain their loyal users over time and find that such communities have smaller, but denser user interaction networks -- with users as nodes, connected if there is a reply-to comment between them. These networks were found to be less assortative and less clustered, and thus show less fragmentation into groups. The authors then predicted whether a user will be loyal to the community, using a machine learning classifier with linguistic features of user's posts and comments, and achieve on average 63.6\% classification accuracy. 

Reddit allows users to self-organize into interest communities, which leads to interesting dynamics of communities. Hessel et al. \cite{Hessel2016} focused on communities sharing name affixes, for example affix ``ask" (science, askscience), ``true" (atheism, trueatheism), ``help" (tech, techhelp), etc. A curious finding is that when such a highly-related community is created, users engaging in the newer community tend to be more active than in their original one. However, in a prevailing number of cases, newer related subreddits do not detach out of their old partners (in terms of the user base), but in about 25\% of the cases, the newer subreddit overtakes its counterpart in participation rate. Some reasons for this behaviour may include the absence of moderation in the new community, a more general scope, or simply a more appealing name spelling. The authors note an interesting result that users who explore the newer created communities generally become relatively more active in their home communities instead of being distracted.

\begin{table}[t]
\begin{tabular}{ L{2.3cm} | C{4.2cm} | C{2.3cm} | R{2.3cm} } 
 Article & Task & Dataset & Methods \\ 
\hhline{====} 
 Glenski et al. \cite{Glenski2017, GlenskiWeninger2017}  & Collect and assess the dataset of tracks of user actions & Reddit, unique dataset of user interactions & Statistical analysis \\ \hline
Singer et al. \cite{Singer2016} & Assess user performance deterioration during activity sessions & Reddit, all comments made in April 2015 & Statistical analysis, negative binomial and Poisson regression \\ \hline
Tan and Lee \cite{Tan2015} & Study explorers and exploring phenomena of new communities & Reddit, dataset \cite{dataset_reddit} & Statistical analysis, regression, linear classification \\ \hline
Hamilton et al. \cite{Hamilton2017} & Loyalty prediction for newcoming users, patterns of loyal communities & Reddit, all comments made in 2014 & User interaction networks, random forest classifiers \\ \hline
Hessel et al. \cite{Hessel2016} & Study the dynamic of arise of highly-related communities & Reddit, dataset \cite{dataset_reddit} & Statistical analysis\\ \hline
Newell et al. \cite{Newell2016} & Study the user migration across platforms during externally caused unrest period & Reddit, dataset \cite{dataset_reddit} & Statistical analysis \\ \hline
Zhang et al. \cite{Zhang2017} & Classify subreddits along ``niche'' and ``volatile'' dimensions, study user retention& Reddit, dataset \cite{dataset_reddit} & Statistical analysis \\ \hline
Das and Lavoie \cite{Das2014effects} & Model users posting strategies with respect to community feedback & Self-collected Reddit dataset & Machine learning, reinforcement learning, Hierarchical Dirichlet Process \\ \hline
Kumar et al. \cite{Kumar2018} & Mobilization and attacks between communities & Reddit, dataset \cite{dataset_reddit} & Reply networks, lexical analysis, LSTM, Mechanical Turk \\ \hline
Tan \cite{Tan2018} & Genealogy of subreddits & Reddit, dataset \cite{dataset_reddit} & Relational networks \\ \hline
\end{tabular} 
\caption{Short summary of the articles with studies on Reddit, presented in Section~\ref{sec:users}}
\end{table}

Users may also migrate under external pressure. In 2015, a series of external events triggered closure of several popular subreddits. Newell et al. \cite{Newell2016} studied the history of this unrest period and observed that users migrating to new subreddits increase their level of participation with respect to their previous community. The authors followed users when they migrated to other discussion platforms and observed that although alternative platforms deliver a space for a broad audience, Reddit users value its advantage of hosting niche communities.  In a similar vein, Zhang et al. \cite{Zhang2017} created a scalable framework for typing subreddits along the ``niche'' and ``volatile'' dimensions and used these types to understand the user retention and assimilation in subreddits. Finally, Muchnik et al. \cite{Muchnik2013} performed a large-scale experiment on a Reddit-like platform to study the herding effect of social influence and how the system reacts to the manipulation of comment scores. They observed that users tend to correct artificially down-voted comments. However, comments that are artificially up-voted received an enhanced number of positive votes, thereby increasing the initial bias. Similar herding effects were found in other social systems as well \cite{Hanson1996, Salganik2008}. Das and Lavoie \cite{Das2014effects} also used a self-collected Reddit dataset of users posts and comments to train a reinforcement-learning model for how users select subreddits to post in reaction to community feedback.

\textbf{Trolling and hate speech.} Chandrasekharan et al. \cite{Chandrasekharan2017} studied the ban of several hate speech subreddits and the consequences that this measure brought to the website. As one could expect the users of the banned subreddits would redistribute themselves over other subreddits and proceed producing hate speech there, but the authors show that did not happen -- the level of hate speech did not increase in other subreddits and, moreover, the majority of users just left the website. Same topic was studied in \cite{Saleem2018}, where the authors note that many counter-actions taken by the users of banned subreddits were short-lived and promptly neutralized by both Reddit moderators. 

In \cite{Chandrasekharan2018} Chandrasekharan et al. studied community norms and their violations. The method comprised continuous scraping of comments and checking their presence in the system 24 hours later, which finally gave more than four millions comments deleted by moderators within a 10-month period. Using state-of-the-art text analysis libraries and principle component analysis the common and particular language norms were inferred, for example, hate speech, racism and homophobia were established to be common norms across the whole Reddit and expressing ``thanks'', mocking religion and nationality were only particular norms, valuable only for a part of subreddits.

Another phenomenon which frequently happens in online discussions is \textit{trolling} -- provocative, offensive or menacing messaging \cite{Bishop2013}. Mojica \cite{Mojica2016} collected and studied an annotated dataset of trolling comments in discussions on Reddit using a variety of language features.

\textbf{Inherent networks of communities.} Kumar et al. \cite{Kumar2018} considered interactions between communities in the form of mobilization by users of a community (the source of the `attack') for hateful comments on posts from another community (the target of the attack). Such mobilization happens when a user in source community posts a link to a post in a target community and titles it with the intention of mobilizing a subset of users, who further write hateful comments on the target post. Such interactions may cause users of a target community to leave. By analyzing reply networks in target discussions, the authors found the effect of echo chambers, i.e. attackers preferentially interact with other attackers and defenders with other defenders. When a direct interaction happen, the attackers ``gang-up'' on defenders and only a small part of the defenders is involved into interactions with the attackers. The authors propose an LSTM neural network model that uses textual and social features in order to identify whether a given cross-linked post will produce a mobilization. 

Gomez et al. \cite{Gomez2008} constructed and analysed the inherent social network of Slashdot, generated by replies in the discussion threads. The network exhibits neutral mixing by degree, almost identical in and out degree distributions, only moderated reciprocity and an absence of a community structure. The authors conjectured that users are more inclined to be linked to people who express different points of view, and that the network may help to identify users with a high diversity in opinions. The authors also proposed a measure for the controversy of a discussion, based on the h-index \cite{Hirsch2005index}. 

Tan \cite{Tan2018} considered the genealogy of communities in Reddit. The author builds a weighted directed network of communities, where community $A$ is linked to a community $B$ if a substantial fraction of first 100 posting users in B had their posts in A. The weight of a link $(A, B)$ is simply the fraction of posting users. The network  shows  user migration across the communities and is useful for predicting growth of communities. One finds that the diverse portfolio of memberships is the most important characteristic of early adopters, whereas community feedback and language similarity does not seem to matter. 

\textbf{Other.} Discussion platforms are a tested for a broad spectrum of possible research questions. In addition to the topics covered above, we give now some other research directions. Derczynski and Rowe \cite{Derczynski2017} used Reddit comments to create an annotated corpus of \textit{named entities} -- proper nouns representing a person, place or an organisation. Horne and Adali \cite{Horne2017Engagement} studied how posting news articles on subreddit \textit{/r/worldnews} influences their popularity and conclude that changing the article titles results in greater popularity comparing to leaving the original one. In a similar way, Moyer et al. \cite{Moyer2015} studied how posts on the subreddit \textit{/r/todayilearned} influenced the pageviews of Wikipedia. 

\section{Discussion}\label{sec:discussion}

This survey does not aim at providing a comprehensive listing of all Reddit-related works, but rather at providing a representative sampling that illustrates the richness of datasets related to online discussion platforms, with Reddit as a dominant example. The richness is clear in terms of quantity as well as quality --- in principle the entire dataset can be harvested, while other platforms, e.g.  Facebook may offer exhaustivity only at the cost of selecting a small sample of volunteers \cite{Lambiotte2014tracking} and studies of Twitter are known to be limited by the volume and bias of their API \cite{morstatter2013sample}. The richness also arises from the diversity of the data, featuring an inherent social network between users, texts constitutive of posts and comments, social appreciation (score), tree structure of posts and comments, all that unfolding in time for years. 

Different platforms may exhibit different type of properties, for instance in the structure of discussion trees, even if they are organised ￼￼￼￼by the same principle. Our discussion on Reddit has shown a broad distribution in size, with a vast majority of trees negligibly small. We also observed root bias, as in
other discussion platforms, however their structure was better modelled as a branching tree with almost uniform branching, rather than preferential attachment networks, due to their specific depth/width profile.

The richness of the data translates in a variety of topics of investigation. On the theoretical level, it allows to observe an ecosystem of users discussing, agreeing or not, and organizing  in communities. Besides fundamental sociological questions, it also allows to investigate a range of Internet-specific questions, such as trolling, echo chambers, polarization, social manipulation,  etc.
The data also offers material to shape solutions for applied questions. From a platform designer viewpoint, it could help to improve the experience of a user, but also to design more efficient algorithms for the identification of high-quality posts. 
The design of the commenting system is also expected to affect the dynamics and structure of conversations. In this direction, important problems include the detection and automatic removal of trolling or attacks, as well as ways to stimulate the activity of a forum.

\begin{figure}[t]
\centering
\includegraphics[width=1\linewidth]{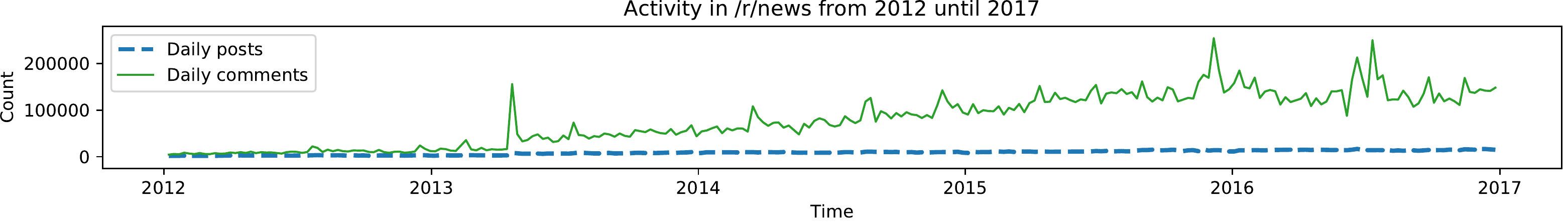}
\includegraphics[width=1\linewidth]{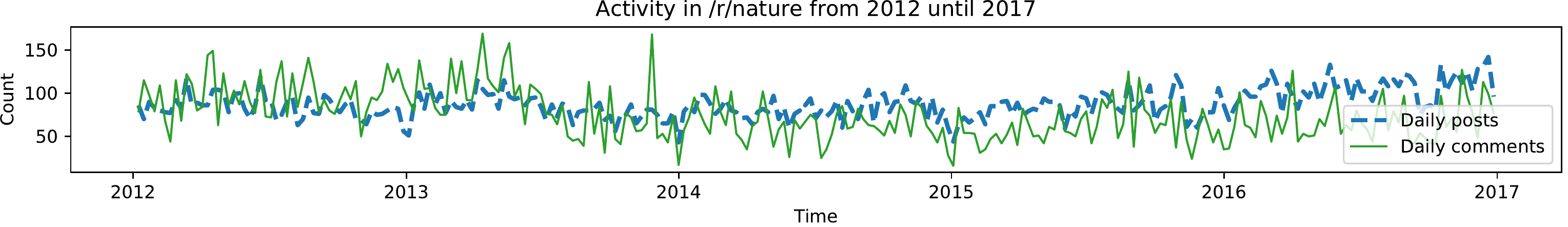}
\includegraphics[width=1\linewidth]{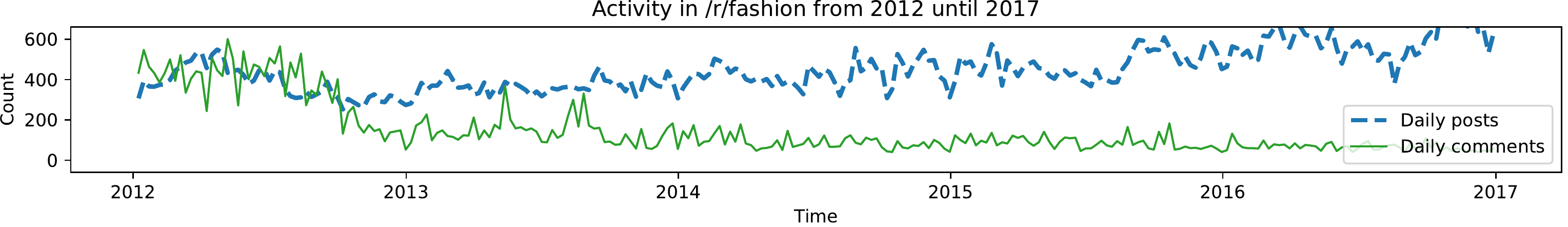}
\caption{Daily counts of submission of posts and comments in three selected subreddits. Three possible scenarios of participation dynamics are shown: 1) increase in comment rate exceeds posting rate (top figure); 2) rates are similar (middle figure); 3) comments eventually disappear, while number of posts increase.}
\label{fig:subreddits_activity}
\end{figure}

The richness of the data and problems calls for a range of computational  methods, which may be explicit statistical models or black-box machine learning tools, in order to classify or predict the behaviors of users, posts, communities. Overall we observe that the structure of discussion trees is relatively well understood. However, mixing the dynamics and structure with textual features is an important step that has only been studied by means of  black box machine learning, such as neural networks techniques, showing a good performance in predicting community appreciation. A challenge remains the fact that many basic statistics of the data (activity of users, popularity of communities, success of a post, etc.) exhibit heavy tails, which may introduce sampling issues as, for instance, a random sampling may fail to observe extreme points (e.g. high activity users) while they carry a large influence in the structure and dynamics of the system. This caveat must be kept in mind when using techniques such as neural networks.

This review is a testimony of the richness and dynamism of academic research on social platforms, in general, and Reddit, in particular. Despite the many progresses overviewed above, we would like to conclude with a list of what we believe to be promising research directions. In our opinion, fruitful avenues of research include:

\begin{itemize}
\item A more detailed study of the activity patterns of users. Collected browsing patterns of users already uncover particular voting behaviour, when many users vote after glancing over the title of a post, and browsing patterns, when feed page breaks create an abrupt obstacle for users attention \cite{Glenski2017, GlenskiWeninger2017}. Activity patterns may first of all be of use to platform designers and to study the influence of platform structure on user experience;

\item It is true that discussion platforms like Reddit do not have an a priori built social network, in comparison
with Twitter or Facebook. Nevertheless, one may reconstruct inherent
networks of communities \cite{Tan2018}, users (as in \cite{Gomez2008}) or submitted information, and
exploit these networks to improve prediction of the
future state or dynamics of the system. Such studies may be of use for platform
users, as well as for platform curators;

\item Study growth or resilience of online communities over time. To be more precise, the dynamics of posting and commenting shape define the life of a community. The two processes are coupled, but not necessarily proportional, as can be seen in Figure \ref{fig:subreddits_activity}. Important questions include the identification of dynamical and structural features that ensure the growth or resilience of online communities over time. Dynamical and evolving graph models would be of help in this direction;

\item The dynamics of discussions is another interesting, yet mostly unexplored, aspect of research, especially the possible relation between the structure and the dynamics of discussion trees. This question could be explored by means of neural networks, which showed to be a working approach in prediction models;

\item Finally, the huge volume of new posts and comments makes the design of efficient ranking and recommendation algorithms vital, in order to allow users to identify relevant information and improve their online experience. As it was shown, platform structure has direct influence on user experience and participation \cite{Aragon2017}, thus both platform designers and users would benefit from such studies.

\end{itemize}

\begin{acknowledgement}
This work was supported by Concerted Research Action (ARC) supported by the Federation Wallonia-Brussels Contract ARC
14/19-060; Flagship European Research Area Network (FLAG-ERA) Joint Transnational Call ``FuturICT 2.0''; and by grant 16-01-00499 of the Russian Foundation for Basic Research. 
\end{acknowledgement}

\bibliographystyle{spmpsci}
\bibliography{literature}

\end{document}